\documentclass[11pt,reqno]{amsart}

\usepackage{graphicx}  
\usepackage{color}  
\usepackage{float}
\usepackage{epstopdf}  

\begin{document} 

\title[Green's function for a sphere]{Green's function for symmetric loading of an elastic sphere with application to contact problems}

\author{A. S. Titovich}
\address{
	Department of Mechanical and Aerospace Engineering\\
	Rutgers University\\
	Piscataway, NJ 08854-8058, USA }
\email{alexey17@eden.rutgers.edu}

\author{A. N. Norris}
\address{
	Department of Mechanical and Aerospace Engineering\\
	Rutgers University\\
	Piscataway, NJ 08854-8058, USA }
\email{norris@rutgers.edu}



\begin{abstract}
A compact form for the static Green's function for symmetric loading of an elastic sphere is  derived. The expression captures the singularity in closed form using standard functions and quickly convergent series. 
Applications to problems involving contact between elastic spheres are discussed. An exact solution for a point load on a sphere is presented and subsequently generalized for distributed loads. Examples for constant and Hertzian-type distributed loads are provided, where the latter is also compared to the Hertz contact theory for identical spheres. The results show that the form of the loading assumed in Hertz contact theory is valid for contact angles up to about 10 degrees. For larger angles, the actual displacement is smaller and the contact surface is no longer flat.
\end{abstract}

\maketitle 


\section{Introduction} \label{sec1}

Contact between spheres has intrigued researchers for more than a century, and still no simple closed-form analytical solution exists. One of the first and most important developments in the field,  due to Heinrich Hertz \cite{Hertz} in 1881, is an  approximate  solution for the normal, frictionless contact of linear elastic spheres.  The major assumption in Hertz's model is that the contact area was small compared to the radii of curvature, which has served as a useful engineering approximation in many applications. Ever since then many have tried to relax this assumption while maintaining a compact, workable solution. The Green's function for symmetric loading on a sphere provides the  means to find the exact response for arbitrary loading, a first step towards improving on Hertz' classic solution. Existing forms of the Green's function are however not suitable for fast and ready computation, either due to slow convergence of series or analytically cumbersome expressions. The goal of the present paper is to provide an alternative form of the  Green's function suitable for fast computation of solutions under arbitrary loading.

Sternberg and Rosenthal \cite{Sternberg52} present an in-depth study of the nature of the singularities on elastic sphere loaded by two opposing concentrated point forces.  As expected,  the dominant inverse square singularity in the stress components can be removed by subtraction of an appropriate multiple of  Boussinesq's solution for a point load at the surface of a half space.  Sternberg and Rosenthal showed that the quickly convergent residual field retains a weaker singularity of logarithmic form, a result that is also evident in the solution developed here. 
The singular solutions obtained by Sternberg and Rosenthal were extended to arbitrarily oriented point forces by Guerrero et al. \cite{Guerrero72}.
Our interest here is in developing an analogous separation of the Green's function (circular ring loading).  In this regard, a relatively compact form of the Green's function for the sphere was derived by Bondareva \cite{Bondareva69} who  used it to solve the problem of the weighted sphere.  In \cite{Bondareva71}, Bondareva formulates an example with a sphere contacting a rigid surface. This has been used to solve the rebound of a sphere from a surface \cite{Villaggio96}.  Bondareva's solution starts with  the known series expansion  \cite{Lurie55} for the  solution of the elasticity problem of a sphere, and replaces it with finite integrals of known functions.  

In this paper we introduce an alternative form for the Green's function for a sphere, comprised of analytical functions and a quickly convergent series. No direct integration is required.  The methodology for determining the analytical functions is motivated by the simple example of a point load on a sphere, for which we derive a solution similar in spirit to that of   \cite{Sternberg52}, but using a fundamentally different approach: partial summation of infinite series as compared with a functional {\it ansatz}.  The present methods allows us to readily generalize the point load solution to arbitrary symmetric normal loading.  A typical contact problem involves solving a complicated integral equation for the contact stress once a displacement is specified. Instead, we will use the derived Green's function in the direct sense, solving for the displacements for a given load. This is used to check the validity of Hertz contact theory through the assumed form of the stress distribution.

The outline of the paper is as follows.   The known series solution for symmetric loading on a sphere is reviewed in 
\S\ref{sec2}.  The proposed method for simplification is first illustrated in \S\ref{sec3} by deriving a quickly convergent form of the solution for a point force.    The Green's function for symmetric loading is then developed in \S\ref{sec4}, and is illustrated by application to different loadings. Conclusions are given in \S\ref{sec5}.

\section{Series solution} \label{sec2}
Consider a solid sphere of radius $R$, with surface $r=R$, $0\leq \theta \leq \pi$, in spherical polar coordinates ($r$, $\theta$, $\phi$). The sphere is linear elastic with shear modulus $G$ and Poisson's ratio $\nu$. The surface is subject to tractions

\begin{equation} \label{-1}
\sigma_{r\theta } = 0, \ \ 
\sigma_{r\phi } = 0, \ \ 
\sigma_{rr} = \sigma (\theta ) , \ \ 
\text{for }\ r=R, \ 0\le \theta \le \pi, \ 0\le \phi < 2\pi .
\end{equation}
Using the known   properties of Legendre functions, see. eqs.\ \eqref{4}, 
allows us to express the normal stress as 
\begin{equation} \label{41}
\sigma (\theta) = \frac 12 \sum\limits_{n=0}^\infty   (2n+1)   \sigma_n P_n(\cos \theta )  , 
\end{equation}
where the Legendre series coefficients are 
\begin{equation} \label{42=1}
\sigma_n =   
 \int\limits_0^\pi  \sigma(\phi) P_n(\cos \phi ) \, \sin  \phi \, d \phi  . 
\end{equation}

The displacements and tractions for the sphere can also be represented in series form  \cite[eq.\ 5]{Zhupanska11}
\begin{equation} \label{1}
\begin{aligned}
2G u_r &= \sum\limits_{n=0}^\infty
\big[
(n-2+4\nu)A_n r + B_nr^{-1} 
\big] \, r^n\, P_n(\cos \theta ) , 
\\
2G u_\theta &= \sum\limits_{n=1}^\infty
\big[
n(n+5-4\nu)A_n r + (n+1) B_nr^{-1} 
\big] \,  r^n\,\frac{ P_n^{1}(\cos \theta ) }{n(n+1)} , 
\\
\sigma_{rr} &= \sum\limits_{n=0}^\infty
\big[
 [ n(n-1) -2(1+\nu)  ]A_n   + (n-1) B_nr^{-2} 
\big] \,  r^n\,P_n(\cos \theta ) , 
\\
\sigma_{r\theta} &= \sum\limits_{n=1}^\infty
\big[
 n [ (n-1)(n+3) +2(1+\nu)  ]A_n   + (n^2-1) B_nr^{-2} 
\big] \,  r^n\, \frac{ P_n^{1}(\cos \theta ) }{n(n+1)} , 
\end{aligned}
\end{equation}
with $B_0 \equiv 0$, and $B_1$ corresponds to a rigid body translation via 
$2G{\bf u} (0,\cdot) = B_1 {\bf e}_z$. 
It follows from \eqref{-1} that 
\begin{equation} \label{2}
\begin{aligned}
&A_0 = \frac{-\sigma_0}{2(1+\nu)} , \ \ A_1 = 0, 
\\ 
&A_n = -\frac{\sigma_n}{4 R^{n}}\, \frac{ (n+1) (2n+1)}{[ n(n-1)+(2n+1)(1+\nu) ]}, \ \ n \ge 2  ,  
\\
& B_n = \frac{-n}{n^2-1}\, [ (n-1)(n+3) +2(1+\nu)  ]\, R^2 A_n,  \ \
n \ge 2.
\end{aligned}
\end{equation}
Thus, noting that  $P_n^1(\cos \theta ) = \frac {d}{d\theta} P_n(\cos \theta )$, we have 
\begin{subequations} \label{6}
\begin{align}
\begin{split}
u_r (R,\theta) 
&= \frac{R}{4 G}\,  \Big( \frac{2(1-2\nu)}{1+\nu} \sigma_0  + 
 \sum\limits_{n=2}^\infty \sigma_n
\bigg(
\frac{2n+1}{n  - 1}
\bigg)
\bigg(
\frac{2(1-\nu) n^2 + \nu n  -1 + 2\nu }{n^2 + (1+2\nu)n +1+\nu} 
\bigg) \,P_n(\cos \theta )  \Big) , 
\\
u_\theta (R,\theta) 
&= \frac{R}{4 G} \, \frac {d}{d\theta}
 \sum\limits_{n=2}^\infty  \sigma_n
\bigg(
\frac{2n+1}{n  - 1}
\bigg)
\bigg(
\frac{(-1+2\nu) n + 2 - \nu }{n^2 + (1+2\nu)n +1+\nu} 
\bigg)  \,P_n(\cos \theta ) .
\end{split}
\end{align}
\end{subequations}

Bondareva \cite{Bondareva69}, using a different representation, replaced the infinite summation of  Legendre functions by a combination of closed form expressions and an integral, each dependent on $\nu$.   The integral term contains a logarithmic singularity which, together with the complex-valued nature of its coefficients, makes its evaluation indirect.   Here we  propose an alternative form for the Green's function in a combination of closed-form expressions and a standard summation of Legendre functions  that is, by design, quickly convergent.

\section{Point force} \label{sec3}

\subsection{Exact solution.}

In order to illustrate the method, we first consider the simpler problem of the point force of magnitude $F$ applied at $\theta = 0$ defined by 
\begin{equation} \label{3}
 \sigma (\theta)= \frac{-F}{2\pi R^2} \, \lim_{\psi \downarrow  0} \, 
\frac{\delta (\theta - \psi)}{\sin \psi} 
\quad
\Leftrightarrow 
\quad
\sigma_n = \frac{-F}{2\pi R^2} 
\end{equation}
where we have used the property  $P_n( 1) = 1$. 
The  difficulty with the infinite summations \eqref{6} is two fold: first,  it is not a suitable form to  reproduce the singular nature of the Green's function;   secondly, it does not  converge quickly as a function of the truncated value for $n$.  The idea here is to replace the summation by closed form expressions plus a summation that is both regular and quickly convergent. The fundamental idea behind the present method is to write $u_r$, $u_\theta$ of eqs.\ \eqref{6} in the form
\begin{subequations} \label{24}
\begin{align} 
u_r (R,\theta)  &= \frac{-F}{8 \pi GR}
\bigg( 4(1-\nu) S(\theta )  + \sum\limits_{j=0}^M a_j (\nu) S_j(\theta )  
+ f(\theta) \bigg),
\\
u_\theta (R,\theta)  &= \frac{-F}{8 \pi GR} \frac {d}{d\theta}
\bigg(  \sum\limits_{j=0}^M b_j (\nu) S_j(\theta )  
+ g(\theta) \bigg), 
\label{4b}
\end{align}
\end{subequations}
where the functions $S(\theta ) $ and $ S_j(\theta )$  $(j=1, \ldots M)$, are closed-form expressions, in this case, 
\begin{subequations} \label{232}
\begin{align}
S (\theta ) & =  \sum\limits_{n=0}^\infty  P_n(\cos \theta )
= \frac 12 \csc \frac \theta 2 ,  \label{232a}
\\
S_j  (\theta )  &=  \sum\limits_{n=0}^\infty   \frac{P_{n+j} (\cos  \theta )}{n+1 }, 
\ \ j =  0, 1, \ldots , 
\end{align}
\end{subequations}
and 
$ f(\theta) $, $ g(\theta) $ 
are regular functions of $\theta$ defined by   quickly convergent series in $n$, 
\begin{equation} \label{25}
\begin{aligned} 
 f(\theta)  =  \sum\limits_{n=0}^\infty  C_n 
   P_n(\cos \theta ) , 
		\quad
	 g(\theta)  =  \sum\limits_{n=0}^\infty  D_n 
   P_n(\cos \theta )	.
\end{aligned}
\end{equation}
The coefficients  $a_0$, $a_1$ $\ldots$  $a_M$ are defined so that  $C_n = $O$(n^{-(M+2)})$ as $n\to \infty$. This criterion  uniquely provides the constants $a_0$, $a_1$ $\ldots$  $a_M$ as solutions of a system of linear equations. Similarly, $b_0$, $b_1$ $\ldots$  $b_M$ are uniquely defined by $D_n = $O$(n^{-(M+2)})$ as $n\to \infty$.

Here we consider the specific case of  $M=2$.  Other values of $M$ could be treated in the same manner; however,  we will show  that $M=2$ is adequate for the purpose of improving convergence. In this case eq.\ \eqref{24} becomes
\begin{subequations} \label{242}
\begin{align} 
u_r (R,\theta)  =& \frac{-F}{ 8\pi GR}\,
\Big[ 4(1-\nu) S(\theta )  +  a_0 S_0(\theta ) +a_1 S_1(\theta )
+a_2 S_2(\theta ) + f(\theta) \Big]
\notag \\
=& \frac{-F}{ 8\pi GR} \, \Big[
\sum\limits_{n=2}^\infty  \Big( 4(1-\nu) + \frac{a_0}{n+1} + \frac{a_1}{n}
+ \frac{a_2}{n-1} + C_n \Big) P_n(\theta )
+ C_0 P_0(\theta)  + C_1 P_1(\theta) 
\notag \\ & + 4(1-\nu)\Big(P_0( \theta ) + P_1(\theta )\Big) + a_0\Big(P_0( \theta ) + \frac{1}{2}P_1(\theta)\Big) + a_1 P_1( \theta )  
 \Big],
\\
u_\theta (R,\theta)  =& \frac{-F}{ 8\pi GR}\, \frac{d}{d\theta}
\Big[ b_0 S_0(\theta ) +b_1 S_1(\theta )
+b_2 S_2(\theta ) + g(\theta) \Big]
\notag \\
=&  \frac{-F}{ 8\pi GR}  \, \frac{d}{d\theta}  \Big[
\sum\limits_{n=2}^\infty \Big( \frac{b_0}{n+1} + \frac{b_1}{n}
+ \frac{b_2}{n-1} + D_n \Big) P_n(\theta )
\notag \\ 
&+  b_0 \Big(P_0( \theta ) + \frac{1}{2}P_1(\theta)\Big) + b_1 P_1(\theta) +  D_0 P_0(\theta)  + D_1 P_1(\theta)  \Big] ,
\end{align} 
\end{subequations}
where the associated three functions  $S_j  (\theta )$, $j=0, 1, 2$ are (see the Appendix)
\begin{subequations} \label{22}
\begin{align} \label{22a}
S_0 (\theta ) & 
= \log \bigg( 1+ \csc \frac \theta 2 \bigg) , 
\\
S_1 (\theta ) & 
= - S_0 (\theta )  - 2\log \sin \frac \theta 2 , 
\label{22b}
\\
S_2 (\theta ) & 
=  S_1 (\theta ) \cos\theta  - 2 \sin \frac \theta 2 \bigg( 1-  \sin \frac \theta 2  \bigg)  . 
\label{22c}
\end{align} 
\end{subequations}

Equations \eqref{24}, \eqref{232a} and \eqref{22} indicate the expected Boussinesq-like $\theta^{-1}$ singularity as well as the weaker $\log  \theta$ singularity first described by Sternberg and Rosenthal \cite{Sternberg52}.   The logarithmic singularities in $S_j  (\theta )$, $j=0, 1, 2$  
can be compared to the potential functions 
$[D_1]$, $[D_2]$, and $[D_3]$ in equation (17) of \cite{Sternberg52}, which provide a  logarithmic singularity.  In the present notation these are, respectively   (using capital $\Phi$s so as not to be confused with the angle $\phi$, and making the substitution $\theta \rightarrow \pi -\theta$) 
\begin{subequations} \label{st1}
\begin{align} \label{rev1}
\Phi_1(\theta ) &= 2 \log  \sin \frac \theta 2  ,
\qquad 
\Phi_2(\theta ) = -R \big( 1+ 2\cos\theta\, \log   \sin \frac \theta 2  \big) ,\\
\Phi_3(\theta ) &= R^2 \big( 2(1 - 3\cos^2\theta) \log  \sin \frac \theta 2   + \cos^2\theta - 3\cos\theta - 1 \big) .
\end{align}
\end{subequations}
These clearly display the same form of the singularity as in equations \eqref{22}, but are otherwise different.

\begin{figure} [H]
\centerline{
\includegraphics[width=4.5in, height=3.2in]{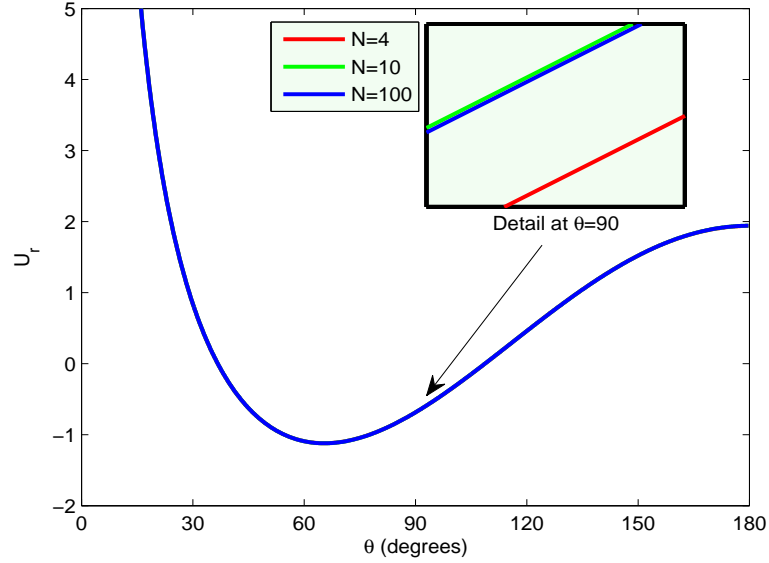}
}
\caption{Convergence of the proposed expression for $U_r = \frac{-8\pi G R}{F} u_r$ in eq.\ \eqref{242}. The magnified view at $\theta = \frac{\pi}{2}$ shows the difference in the value of $U_r$, which is $(U_r)_{N=10}-(U_r)_{N=4}=4.5259\cdot 10^{-4}$ and $(U_r)_{N=100}-(U_r)_{N=10}=-1.7059\cdot 10^{-5}$. The small plot is magnified by a factor of 11250. }
\label{Fig1}
\end{figure}

Define the first two coefficients of $f(\theta)$ and $g(\theta)$ from \eqref{25} as
\begin{subequations} \label{23}
\begin{align}
C_0 &= \frac{2(1-2\nu)}{1+\nu} - 4(1-\nu) - a_0 ,
\quad
C_1 = - 4(1-\nu) - \frac{1}{2}a_0 - a_1 , 
\\ 
D_0 &= - b_0 ,
\qquad\qquad\qquad\qquad\qquad
D_1 = - \frac{1}{2}b_0 -  b_1 . 
\end{align} 
\end{subequations}
The coefficients $a_n$ and $b_n$ are then found by comparing expression \eqref{242} to the series solution in \eqref{6}, expanding both expressions for large $n$, and equating the coefficients of the same order terms.  Thus, the original  assumed form of the solution \eqref{24} implies
\begin{equation} \label{7}
\begin{aligned}
& 
\sum\limits_{n=2}^\infty 
\frac{P_n(\theta )   }{n^2 + (1+2\nu)n +1+\nu} 
\Big(
\frac{2n+1}{n - 1}
\Big)
\times
\begin{cases}
\big(
2(1-\nu) n^2 + \nu n  -1 + 2\nu 
\big) 
\\
\big(
(-1+2\nu) n + 2 - \nu
\big) 
\end{cases}
\\
&\qquad \qquad \qquad \qquad \qquad 
=  \sum\limits_{n=2}^\infty   
 P_n(\theta ) \times
 \begin{cases}
\big( 4(1-\nu) + \frac{a_0}{n+1} + \frac{a_1}{n}
+ \frac{a_2}{n-1} + C_n  \big)  , 
\\
 \big(  \frac{b_0}{n+1} + \frac{b_1}{n}
+ \frac{b_2}{n-1} + D_n \big),
\end{cases}
\end{aligned}
\end{equation}
where 
\begin{subequations} \label{26}
\begin{align}
\begin{split}
 a_0 & = 
\frac 12 (1+\nu) (1-2\nu) (-16\nu^2 + 8\nu + 5) 
, 
 \\
 a_1 & = -32\nu^4 + 16\nu^3 + 30\nu^2 - 16\nu -1, 
  \\
 a_2 & = 16\nu^4 - 16\nu^3-5\nu^2 + \frac{13}{2}\nu + \frac{1}{2}, 
  \\
 b_0 & = 
 \frac 12 (1+\nu) (16\nu^2 - 12\nu - 1) 
 , 
  \\
  b_1 & = -16\nu^3 + 4\nu^2 + 13\nu - 4, 
   \\
  b_2 & = 8\nu^3 - 6\nu^2 - \frac{5}{2}\nu + \frac{5}{2} .
	\end{split}
\end{align} 
\end{subequations}
The remaining coefficients $C_n$ and $D_n$ are then determined directly from \eqref{7}
\begin{subequations} \label{251}
\begin{align}
C_n &= \frac{(1+\nu)}{L_n} \Big( (6-a_1-6a_2)n +a_1  \Big)
\notag \\
&= -\frac{(1+\nu)}{L_n} \Big(
(64\nu^4 - 80\nu^3 +23\nu - 4)n +(32\nu^4 - 16\nu^3 - 30\nu^2 + 16\nu + 1)
\Big)   ,
  \\
D_n &= \frac{(1+\nu)}{L_n} \Big( (6-b_1-6b_2)n + b_1  \Big)
\notag \\
&= -\frac{(1+\nu)}{L_n} \Big(
(32\nu^3 - 32\nu^2 - 2\nu + 5)n +(16\nu^3 - 4\nu^2 - 13\nu + 4)
\big) ,
\end{align} 
\end{subequations}
where 
\begin{equation} \label{3-3-3}
L_n \equiv n(n^2-1)(n^2 + (1+2\nu)n +1+\nu) .
\end{equation}
In summary, the new form of the point force solution is given by the displacements in eq.\ \eqref{242} where the functions and coefficients are given in eqs.\ \eqref{22}-\eqref{23}, \eqref{26}-\eqref{3-3-3}.

It is useful to note that the present approach, like that of Sternberg and Rosenthal \cite{Sternberg52}, leads to a separation of the solution into a part that displays the singular behavior plus an additional part that is quickly convergent.    Sternberg and Rosenthal's solution was based on an {\it ansatz} \cite[eq.\ (38) ]{Sternberg52}  motivated by separation of the stress  into a Boussinesq term with leading order singularity plus a residual field.  Our approach is somewhat different, in that we regularize the infinite series solution directly by partial summation.  The net effect is essentially the same as in \cite[eq.\ (38) ]{Sternberg52} for the point load.   However, the present method can be easily adapted to the more general case of the Green's function for a circular (ring) load, see \S\ref{sec4}.   The analogous generalization of Sternberg and Rosenthal's method is not as obvious.

\subsection{Numerical examples.}

\begin{figure} [H]
\centerline{
\includegraphics[width=4.5in, height=3.2in]{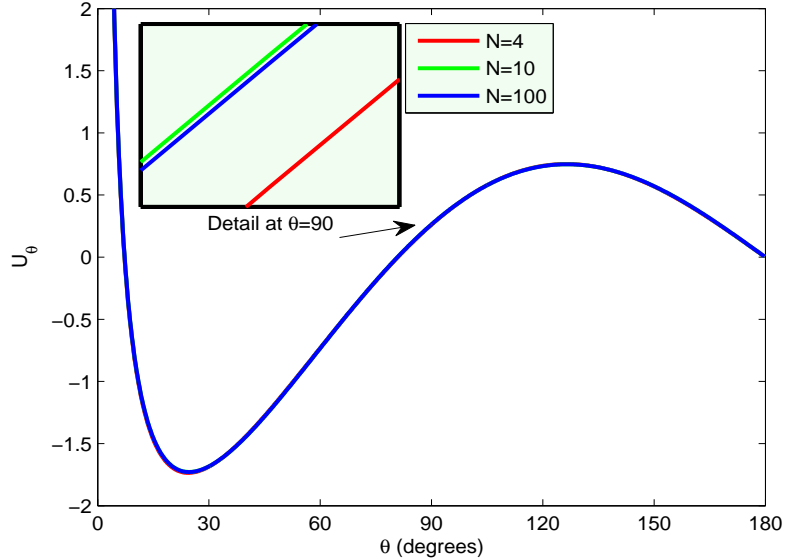}
}
\caption{Convergence of $U_\theta = \frac{-8\pi G R}{F} u_\theta$ given in eq.\ \eqref{242}. The magnified view at $\theta = \frac{\pi}{2}$ shows the difference in the value of $U_\theta$, which is $(U_\theta)_{N=10}-(U_\theta)_{N=4}=2.4097\cdot 10^{-3}$ and $(U_\theta)_{N=100}-(U_\theta)_{N=10}=-1.5040\cdot 10^{-4}$. The small plot is magnified by a factor of 1200. } 
\label{Fig2}
\end{figure}

\begin{figure} [!h]
\centerline{
\includegraphics[width=6.5in, height=2.in]{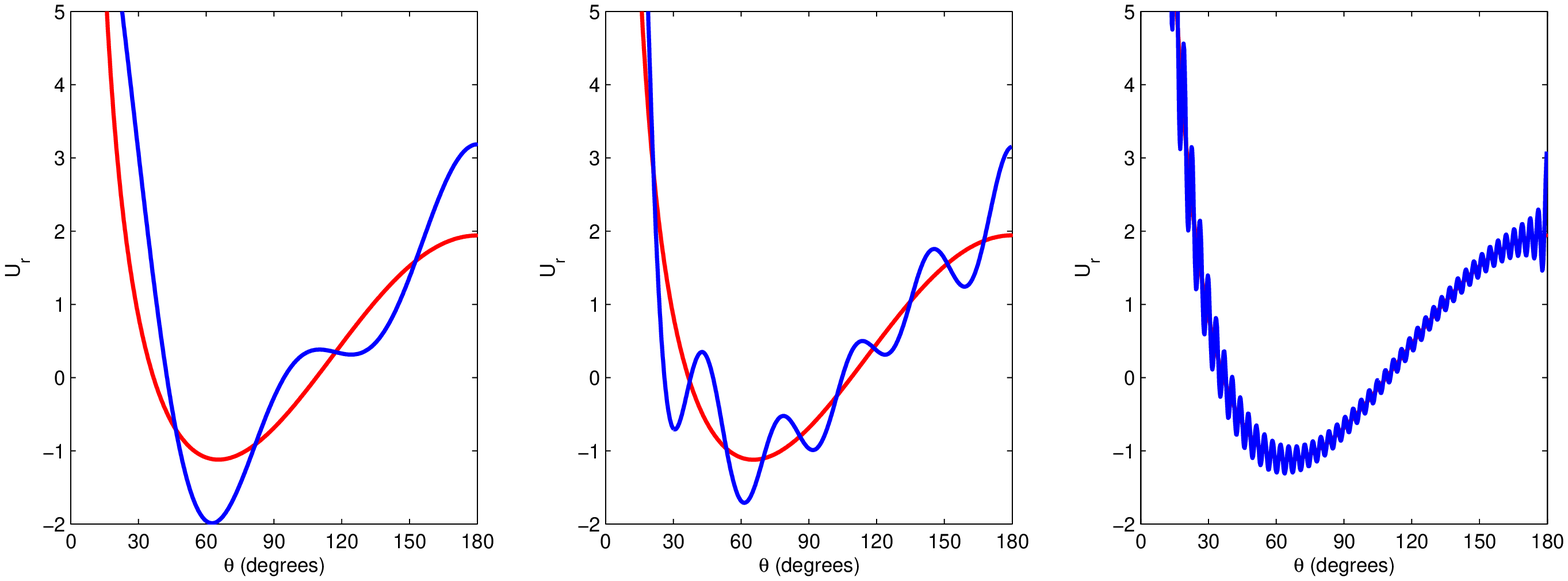}
}
\caption{Comparison of the convergence of $U_r = \frac{-8\pi G R}{F} u_r$ with the truncation value $N$ for the existing expression (eq.\ \eqref{6}, shown in blue) and the expression proposed herein (eq.\ \eqref{242}, shown in red). Left to right: $N$=4, 10, 100.}
\label{Fig3}
\end{figure}

\begin{figure} [!h]
\centerline{
\includegraphics[width=6.5in, height=2.in]{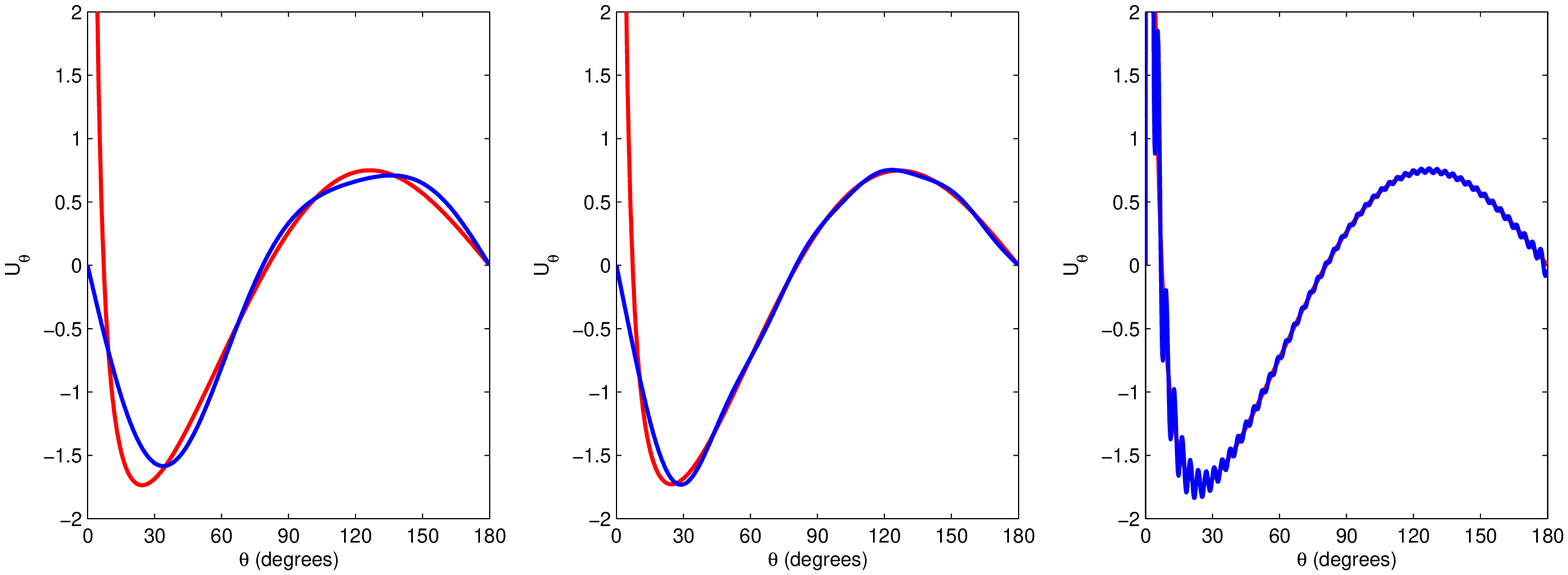}
}
\caption{Comparison of the convergence of $U_{\theta} = \frac{-8\pi G R}{F} u_{\theta}$ with the truncation value $N$ for the existing expression (eq.\ \eqref{6}, shown in blue) and the expression proposed herein (eq.\ \eqref{242}, shown in red). Left to right: $N$=4, 10, 100.}
\label{Fig4}
\end{figure}

In the following examples we introduce the integer $N$ as the truncation value of the series in eqs.\ \eqref{25}. The Poisson's ratio was taken to be $0.4$. Displacements have been normalized by the constant coefficient of the series as $U_i =  {-8\pi G R}F^{-1} u_i$, where $ i=r,\theta$. Figures \ref{Fig1} and \ref{Fig2} show the rate of convergence of the displacements given by eq.\ \eqref{242}, whereas Figures \ref{Fig3} and \ref{Fig4} compare the displacements in eqs.\ \eqref{6} with eqs.\ \eqref{242}.

By design, the proposed expression (eq.\ \eqref{242}) converges much faster than the existing expression (eq.\ \eqref{6}) as seen in Figures \ref{Fig3} and \ref{Fig4}. Looking at the convergence of the proposed expressions with the truncation value $N$, Figures \ref{Fig1} and \ref{Fig2}, we can suggest that the analytic portion of the expression alone gives close results. However, it should be noted that one cannot get rid of the first two terms in the series for $f(\theta)$ and $g(\theta)$ because of their large magnitudes. As far as the general behaviour of the normalized displacements with $\theta$, we see that they increase asymptotically approaching $\theta=0$, change sign between $36.7^\circ$ and $108.7^\circ$ for $U_r$ ($7.27^\circ$ and $80.83^\circ$ for $U_\theta$), and have a minimum at $65.5^\circ$ for $U_r$ ($24.6^\circ$ for $U_\theta$). This is difficult to see in the Figures, but due to symmetry of the loading, the displacement $U_\theta$ must have a value of zero at $\theta=0$.

\section{Green's function}  \label{sec4}

\subsection{A fast convergent form for the Green's function}

The surface displacements for arbitrary loading may be written, by analogy with the  
{\em ansatz} \eqref{24} for the point force, and generalizing the latter, 
\begin{subequations} \label{34}
\begin{align}
u_r (R,\theta)  &= \frac{R}{4 G} \int_0^\pi 
\bigg(4(1-\nu) S(\theta ,\phi )  + \sum\limits_{j=0}^M a_j (\nu) S_j(\theta ,\phi)  
+ f(\theta ,\phi) \bigg) \, \sigma (\phi) \sin\phi d \phi , 
\\
u_\theta (R,\theta)  &=\frac{R}{4 G} \frac {d}{d\theta} \int_0^\pi 
\bigg(  \sum\limits_{j=0}^M b_j (\nu) S_j(\theta ,\phi)  
+ g(\theta ,\phi) \bigg)\, \sigma (\phi) \sin\phi d \phi  , 
\end{align} 
\end{subequations}
where $S(\theta ,\phi) $ and $ S_j(\theta ,\phi)$  $(j=1, \ldots M)$, are   
\begin{subequations} \label{332}
\begin{align}
S (\theta ,\phi) & =  \sum\limits_{n=0}^\infty  P_n(\cos \theta )P_n(\cos \phi ), 
\\
S_j  (\theta ,\phi)  &=  \sum\limits_{n=0}^\infty   \frac{1}{n+1 } P_{n+j} (\cos  \theta ) P_{n+j} (\cos  \phi ), 
\ \ j =  0, 1, \ldots  ,
\end{align} 
\end{subequations}
and 
$ f(\theta ,\phi) $, $ g(\theta ,\phi) $
are regular functions of $\theta$ defined by   quickly convergent series in $n$, 
\begin{align}
 f(\theta ,\phi)  =  \sum\limits_{n=0}^\infty  C_n 
   P_n(\cos \theta ) P_n(\cos \phi ), 
		\quad
	 g(\theta ,\phi)  =  \sum\limits_{n=0}^\infty  D_n 
   P_n(\cos \theta )	P_n(\cos \phi ).
\end{align}
The coefficients  $a_0$, $a_1$ $\ldots$  $a_M$ are the same as before.   The main complication is to find the functions \eqref{332}.   Thus, $S (\theta ,\phi)$ follows from  \eqref{1-4} as 
\begin{equation} \label{-2}
S (\theta ,\phi )  =  
\begin{cases}
\frac 1{\pi} \csc \frac {\theta}2 \sec \frac {\phi}2 \, 
K\Big(\big( \cot \frac {\theta}2 \tan \frac {\phi}2
\big)^2 \Big), & \theta > \phi, 
\\
\frac 1{\pi} \sec \frac {\theta}2 \csc \frac {\phi}2 \, 
K\Big( \big( \tan \frac {\theta}2 \cot \frac {\phi}2
\big)^2\Big), & \theta < \phi ,
\end{cases}
\end{equation}
where $K(m)$ is the complete elliptic integral of the first kind \cite[17.3.1]{Abramowitz74}, while   eq.\  \eqref{1-2} implies 
\begin{equation} \label{3-2}
S_{01}(\theta ,\phi) \equiv 
 S_0(\theta,\phi)+S_1(\theta,\phi) = \begin{cases}
- 2\log  	\sin \frac {\theta}2 \cos \frac {\phi}2  , & \theta > \phi, 
\\
- 2\log  	\cos \frac {\theta}2 \sin \frac {\phi}2  , & \theta < \phi .
\end{cases} 
\end{equation}
The  functions $S_j  (\theta ,\phi) $ can be determined, but their form is overly complicated, and defeats our objective  of simplifying the Green's function.  We therefore restrict the solution to the use of the above two series: $S(\theta ,\phi)$ and $S_{01}(\theta ,\phi)$.

We therefore consider the following form of the {\em ansatz} \eqref{34} using the series $S$ and $S_{01}$  of eqs.\ \eqref{-2} and \eqref{3-2}, respectively.  Substituting them into  eqs.\ \eqref{34} yields the identities 
\begin{subequations} \label{40}
\begin{align} \label{3+=}
u_r (R,\theta)  =& \frac{R}{4 G} \int_0^\pi 
\bigg( 4(1-\nu) S(\theta ,\phi )  + a_{01}(\nu) S_{01}(\theta ,\phi)  
+ f(\theta ,\phi) \bigg) \, \sigma (\phi) \sin\phi d \phi , 
\notag \\
=& \frac{R}{4 G} \int_0^\pi \bigg[ \sum\limits_{n=2}^\infty \bigg( 4(1-\nu)  + a_{01} \bigg(\frac{2n+1}{n(n+1)}\bigg) + C_n \bigg) P_n(\cos\theta)P_n(\cos\phi) 
\notag \\
&+ 4(1-\nu) \bigg( P_0(\cos\theta)P_0(\cos\phi) + P_1(\cos\theta)P_1(\cos\phi) \bigg)
\notag \\
&+ a_{01} \bigg( P_0(\cos\theta)P_0(\cos\phi) + \frac{3}{2}P_1(\cos\theta)P_1(\cos\phi) \bigg)
\notag \\
&+ C_0 P_0(\cos\theta)P_0(\cos\phi) + C_1 P_1(\cos\theta)P_1(\cos\phi) \bigg] \, \sigma (\phi) \sin\phi d \phi ,
\\
u_\theta (R,\theta)  =& \frac{R}{4 G} \frac {d}{d\theta} \int_0^\pi 
\bigg( b_{01} (\nu) S_{01}(\theta ,\phi)  
+ g(\theta ,\phi) \bigg)\, \sigma (\phi) \sin\phi d \phi  ,
\notag \\
=& \frac{R}{4 G} \frac {d}{d\theta} \int_0^\pi \bigg[  \sum\limits_{n=2}^\infty  \bigg( b_{01} \bigg(\frac{2n+1}{n(n+1)}\bigg) + D_n  \bigg) P_n(\cos\theta)P_n(\cos\phi) 
\notag \\
&+ b_{01} \bigg( P_0(\cos\theta)P_0(\cos\phi) + \frac{3}{2}P_1(\cos\theta)P_1(\cos\phi) \bigg)
\notag \\
&+  D_0 P_0(\cos\theta)P_0(\cos\phi) + D_1 P_1(\cos\theta)P_1(\cos\phi)  \bigg] \, \sigma (\phi) \sin\phi d \phi .
\end{align} 
\end{subequations}
Once again we define the first two coefficients of $f(\theta ,\phi)$ and $g(\theta ,\phi)$ as
\begin{subequations} \label{41=}
\begin{align}
C_0 &= \frac{2(1-2\nu)}{1+\nu} - 4(1-\nu) - a_{01} ,
\quad 
C_1 = - 4(1-\nu) - \frac{3}{2}a_{01} , 
\\ 
D_0 &= - b_{01} ,
\qquad \qquad \qquad 
D_1  = - \frac{3}{2}b_{01} ,
\end{align} 
\end{subequations}
which allows us to solve the following expressions for the coefficients $a_{01}$ and $b_{01}$.
\begin{subequations} \label{42}
\begin{align}
& \sum\limits_{n=2}^\infty 
\Big(
\frac{2n+1}{n - 1}
\Big)
\Big(
\frac{2(1-\nu) n^2 + \nu n  -1 + 2\nu }{n^2 + (1+2\nu)n +1+\nu} \Big) P_n(\cos\theta)P_n(\cos\phi)
\notag \\
& \qquad \qquad  \qquad  
= \sum\limits_{n=2}^\infty \Big( 4(1-\nu) +  \frac{(2n+1)}{n(n+1)}  a_{01} + C_n \Big)P_n(\cos\theta)P_n(\cos\phi), 
\\
&\sum\limits_{n=2}^\infty  \Big(
\frac{2n+1}{n  - 1}
\Big)
\Big(
\frac{(-1+2\nu) n + 2 - \nu }{n^2 + (1+2\nu)n +1+\nu}  \bigg) P_n(\cos\theta)P_n(\cos\phi)
\notag \\
&\qquad \qquad  \qquad  
  =  \sum\limits_{n=2}^\infty \bigg(  \frac{(2n+1)}{n(n+1)} b_{01}  + D_n  \Big) P_n(\cos\theta)P_n(\cos\phi).
\end{align} 
\end{subequations}
This is done by expanding equations \eqref{42} for large $n$ and equating same order terms yielding
\begin{equation} \label{43}
 a_{01}   = (2\nu - 1)^2 ,
\qquad
 b_{01}  = 2\nu - 1 .
\end{equation}
Using \eqref{43}, $C_n$ and $D_n$ are found directly from equation \eqref{42} (see also \eqref{3-3-3})
\begin{subequations} \label{44}
\begin{align}
C_n =& 
-  \frac{1}{L_n} \big[
(\nu-1)(4\nu-1)(4\nu+1)n^3 + (8\nu^2 - 11\nu - 1 )n^2 
\notag \\
& \qquad 
+ (-12\nu^3 + 8\nu^2 + 3\nu - 5)n - (\nu+1)(2\nu-1)^2 
\big]  ,
\\
D_n =&  
- \frac{(2n+1)}{L_n} \big[ (\nu - 1)(4\nu+1)n^2 + 2(-\nu^2 + \nu - 1)n - (\nu + 1)(2\nu-1)\big] . 
\end{align} 
\end{subequations}

In summary, 
\begin{subequations} \label{4=5}
\begin{align} \label{4=6}
u_i (R,\theta)  =& \frac{R}{4 G} \int_0^\pi 
H_i(\theta ,\phi) 
\, \sigma (\phi) \sin\phi d \phi , \ \ i = r,\theta ,
 \\
H_r(\theta ,\phi)  =& 
 4(1-\nu) S(\theta ,\phi )  + (1-2\nu )^2  S_{01}(\theta ,\phi)  
+ 
 \sum\limits_{n=0}^\infty  C_n 
   P_n(\cos \theta ) P_n(\cos \phi ), 
\\
H_\theta(\theta ,\phi)  =& 
 \frac {d}{d\theta} 
\bigg( (2\nu - 1)  S_{01}(\theta ,\phi)  
+  \sum\limits_{n=0}^\infty  D_n 
   P_n(\cos \theta ) P_n(\cos \phi )  \bigg)  
\end{align} 
\end{subequations}
where the coefficients $C_n$, $D_n$ are given in \eqref{44}.  
Note that $C_n,D_n=$O$(n^{-2})$ as $n\to \infty$, ensuring   rapidly convergent series. 
The Green's functions of \eqref{4=5} are generally valid for $\theta \in [0,\pi]$.   The 
integrands $H_i(\theta ,\phi) $ are smooth  and bounded functions of $\phi$ for $\phi \ne \theta$, which is always the case if the displacements are evaluated at points outside the region of the loading $\sigma (\phi)$.  However, for points under the load, the integration of $H_r(\theta ,\phi) $ involves a  logarithmic singularity at $\phi = \theta$.   A simple means of dealing with this is described next.

\begin{figure} [htb]
\centerline{
\includegraphics[width=6.5in]{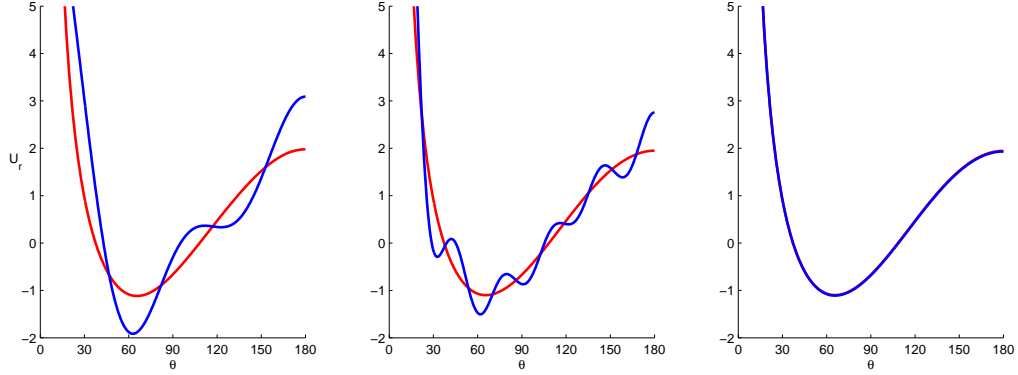}
}
\caption{Comparison of the proposed solution (eq. \eqref{333} shown in red) and existing series solutions (eq. \eqref{6} shown in blue) for $U_r = \frac{-8\pi G R}{F} u_r$ under a Hertzian-type load distributed up to $\phi=10^\circ$. Left to right: N=4, 10, 100.}
\label{FigHSr}
\end{figure}
\begin{figure} [htb]
\centerline{
\includegraphics[width=6.5in]{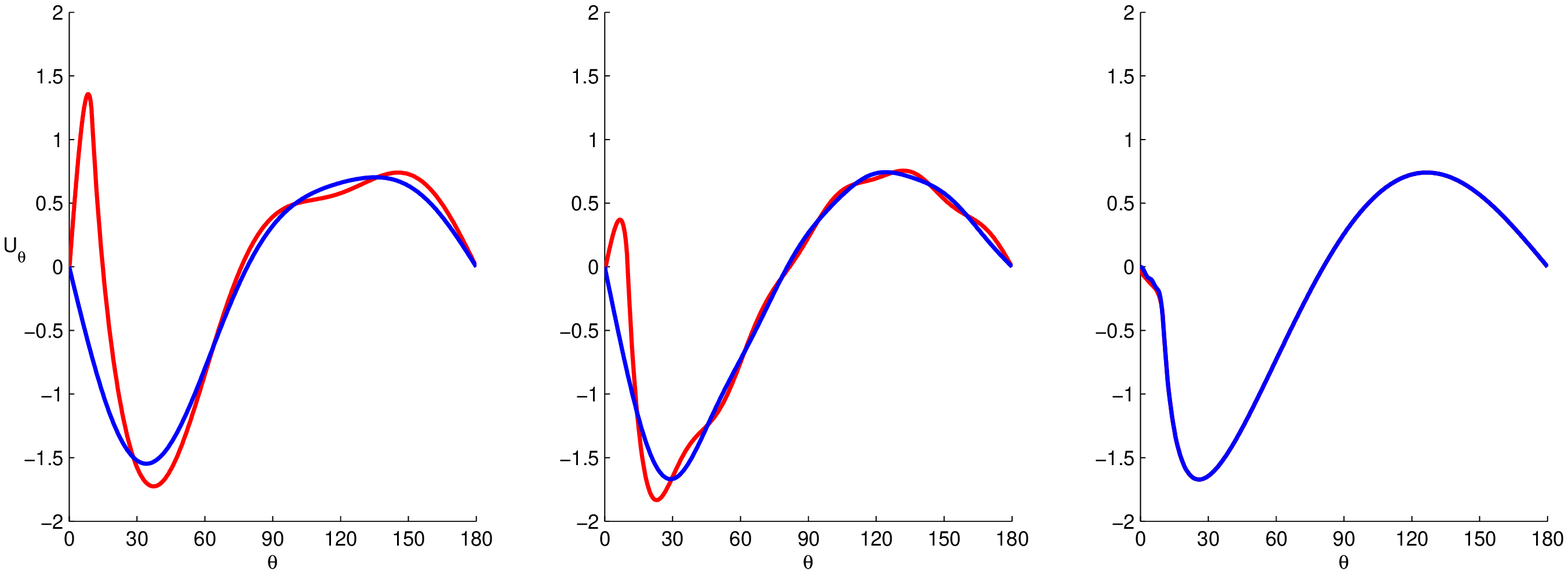}
}
\caption{Comparison of the proposed solution (eq. \eqref{333} shown in red) and existing series solutions (eq. \eqref{6} shown in blue) for $U_\theta = \frac{-8\pi G R}{F} u_\theta$ under a Hertzian-type load distributed up to $\phi=10^\circ$. Left to right: N=4, 10, 100.}
\label{FigHSt}
\end{figure}

\begin{figure} [htb]
\centerline{
\includegraphics[width=6.5in]{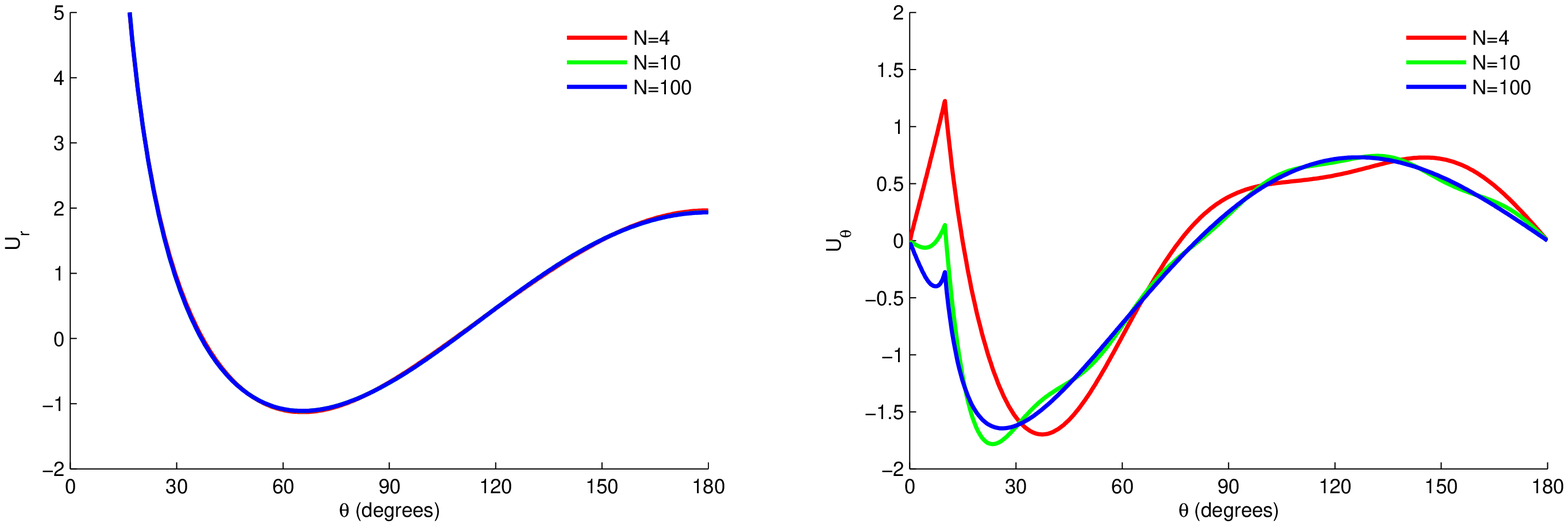}
}
\caption{Convergence of the expressions for $U_r = \frac{-8\pi G R}{F} u_r$ and $U_\theta = \frac{-8\pi G R}{F} u_\theta$ given in equation \eqref{40} with $N$ for a constant distributed load given by \eqref{45}. The load was distributed up to $\phi_0=10^\circ$.}
\label{Fig5}
\end{figure}

\subsubsection{Removing the singularity under the load}

The function $S(\theta,\phi)$ exhibits a logarithmic singularity by virtue of the 
asymptotic behavior 
\begin{equation} \label{34-}
K(m) = \log \frac{4}{\sqrt{1-m}} + \text{O} (1-m), \ \ \ m\uparrow 1.
\end{equation}
The integral in \eqref{3+=} is evaluated by rewriting eq.\ \eqref{3+=}  in the equivalent form
\begin{align} \label{844}
u_r (R,\theta)  =& \frac{R}{4 G} \Big\{ \int_0^{\phi_0}
\Big[
\big( a_{01}(\nu) S_{01}(\theta ,\phi)  
+ f(\theta ,\phi) \big) \, \sigma (\phi) 
\notag \\
 & +   4(1-\nu) \big( S(\theta ,\phi )\sigma (\phi)   
 - \hat S (\theta ,\phi )\sigma (\theta)   \big)
 \Big]
 \sin\phi d \phi 
 \notag \\
 &+  4(1-\nu) \sigma (\theta)  \int_0^{\phi_0}  \hat S (\theta ,\phi )\sin\phi d \phi 
 \Big\}
 ,  \ \ 0 \le \theta \le \phi_0,
\end{align}
where the angle $\phi_0$ defines the domain of the loading, which is normally for contact problems, much less that $\pi$. The function $\hat S (\theta ,\phi )$ has the same singularity as $ S (\theta ,\phi )$ and has a relatively simple integral.  We choose
\begin{equation} \label{-244}
\hat S (\theta ,\phi )  =  
\begin{cases} -
\frac 1{2\pi} \csc \frac {\theta}2 \sec \frac {\phi}2 \, 
\log (\cos^2 \tfrac\phi 2 -\cos^2 \tfrac\theta 2 )  , & \theta > \phi, 
\\
-
\frac 1{2\pi} \sec \frac {\theta}2 \csc \frac {\phi}2 \, 
\log (\sin^2 \tfrac\phi 2 -\sin^2 \tfrac\theta 2 )  , & \theta < \phi ,
\end{cases}
\end{equation}
The integrand of  the first integral in \eqref{844} is now a smoothly varying function with no singularity, and the  second integral is, explicitly, 
\begin{align} \label{825}
  \int_0^{\phi_0}  \hat S (\theta ,\phi )\sin\phi d \phi =&
 - \frac{2}{\pi \sin\frac\theta 2}
  \int_{\cos\frac\theta 2}^1 
  \log (x^2 - \cos^2\tfrac\theta 2 ) d x
  - \frac{2}{\pi \cos\frac\theta 2}
  \int_{\sin\frac\theta 2}^{\sin\frac{\phi_0}{2}}
  \log (x^2 - \sin^2\tfrac\theta 2 ) d x
  \notag \\
  =& \frac{G(\cos\tfrac\theta 2 ,1)}{ \sin\frac\theta 2} 
   + 
  \frac{ G(\sin\tfrac\theta 2 , \sin\tfrac{\phi_0}{2} )}{  \cos\frac\theta 2}  ,
  \ \ 0 \le \theta \le \phi_0,  \ \quad  \text{where} 
  \\
  G(x,y)=& - \frac{2}{\pi} \big( (y-x)\log (y-x)+(y+x)\log (y+x)
  - 2(y - x + x \log  2x )  
  \big)
 .
\end{align}
In summary, the solution for $u_r$ with the singularity removed has the following form (see also eq.\ \eqref{4=6} for $H_r (\theta ,\phi )$ and eq.\ \eqref{825} for $G(x,y)$)
\begin{equation} \label{333}
\begin{aligned}
u_r (R,\theta)  =& \frac{R}{4 G} \Big\{ \int_0^{\phi_0} \bigg[ H_r (\theta ,\phi ) \sigma (\phi) - \hat H_r (\theta ,\phi ) \sigma (\theta) \bigg] \sin\phi d \phi 
+  h(\theta) \Big\} ,
 \\
 \hat H_r (\theta ,\phi ) =& 4(1-\nu) \hat S (\theta ,\phi ) ,
 \\
 h(\theta) =& 4(1-\nu) \bigg[ \frac{G(\cos\tfrac\theta 2 ,1)}{ \sin\frac\theta 2} + \frac{ G(\sin\tfrac\theta 2 , \sin\tfrac{\phi_0}{2} )}{  \cos\frac\theta 2} \bigg]\sigma (\theta)  .
 \end{aligned}
\end{equation}

\begin{figure} [h]
\centerline{
\includegraphics[width=6.5in]{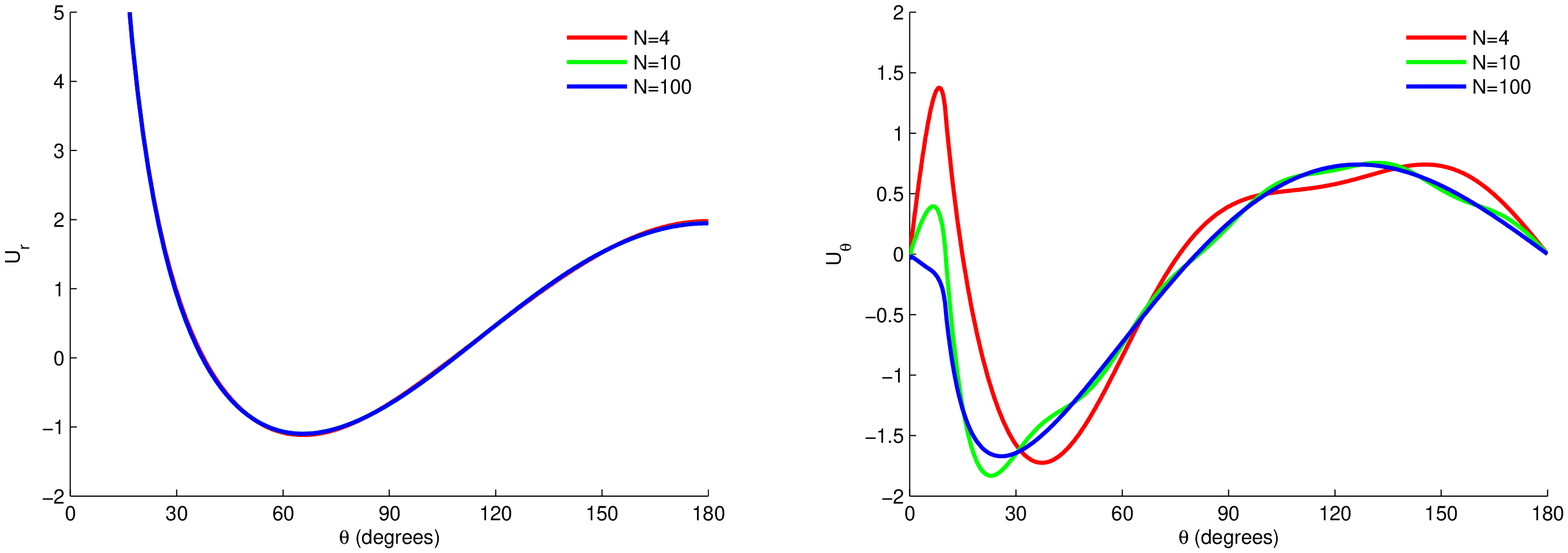}
}
\caption{Convergence of the expressions for $U_r = \frac{-8\pi G R}{F} u_r$ and $U_\theta = \frac{-8\pi G R}{F} u_\theta$ given in equation \eqref{40} with $N$ for a Hertzian-type distributed load given by \eqref{46}. The load was distributed up to $\phi_0=10^\circ$.} 
\label{Fig6}
\end{figure}

\subsection{Examples of distributed loads}

To check the convergence of the expressions in \eqref{40} we will consider a symmetric constant distributed load $\sigma(\phi)$ of the form
\begin{equation} \label{45}
\sigma(\phi) = \frac{-F}{\pi R^2} \frac{1}{ \sin^2\phi_0} , \;\;\;  0 \leq \phi \leq \phi_0 ,
\end{equation}
and a symmetric Hertzian-type load of the form
\begin{equation} \label{46}
\sigma(\phi) = \frac{-3F}{2\pi R^2} \frac{\sqrt{\sin^2\phi_0 - \sin^2\phi}}{ \sin^3\phi_0} , \;\;\;  0 \leq \phi \leq \phi_0 .
\end{equation}
Both loads have been normalized such that their resultant forces are $-F$ for all ranges of the angle $\phi_0$, which is equivalent to the point force given by equation \eqref{3}. The solution on the interval $0\le \theta\le \phi_0$ is obtained using \eqref{333} and for $\phi_0<\theta\le \pi$ we apply equations \eqref{4=5} directly.

\begin{figure} [h]
\centerline{
\includegraphics[width=6.5in]{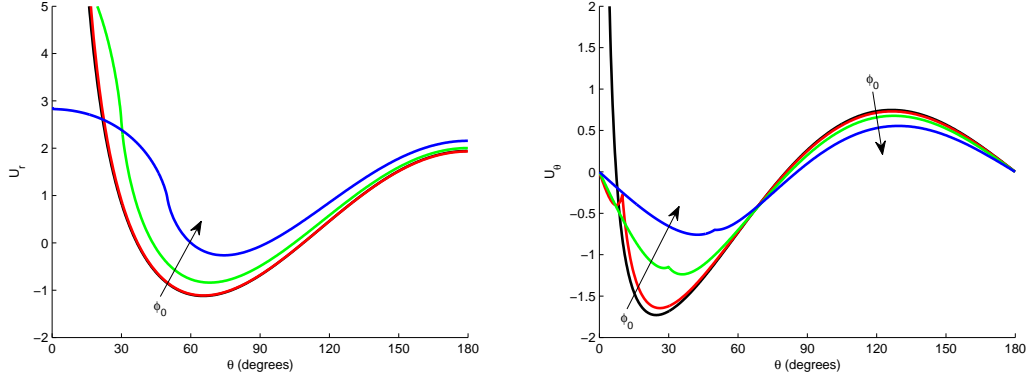}
}
\caption{Comparison of the displacement due to a constant distributed load to the displacement due to a point force of the same magnitude (black). The solution was truncated at $N$=300. The loads were distributed over $\phi_0$=$10^\circ$(red), $30^\circ$(green), $50^\circ$(blue).}
\label{Fig7}
\end{figure}

\begin{figure} [h]
\centerline{
\includegraphics[width=6.5in]{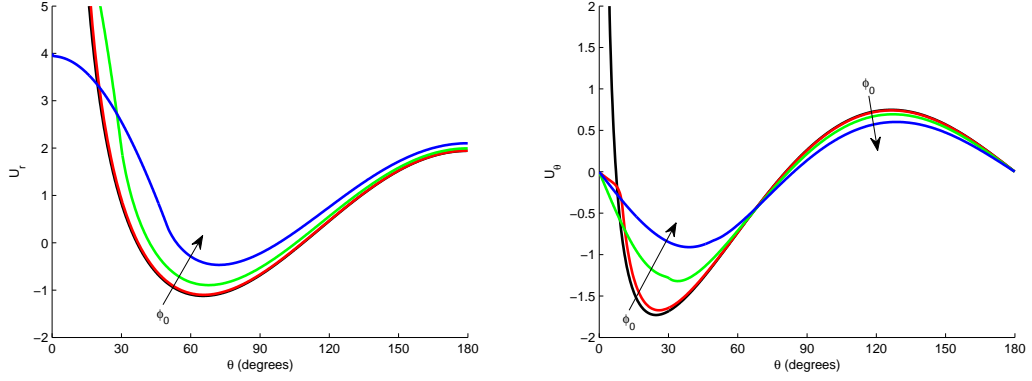}
}
\caption{Convergence of the displacement due to a Hertzian-type distributed load to the displacement due to a point force of the same magnitude (black). The solution was truncated at $N$=300. The loads were distributed over $\phi_0$=$10^\circ$(red), $30^\circ$(green), $50^\circ$(blue).} 
\label{Fig8}
\end{figure}

Firstly, the convergence of the proposed solution (eq. \eqref{333}) is compared to the series solution for a Hertzian-type load in Figures \ref{FigHSr} and \ref{FigHSt}.  These curves indicate that  the convergence of the radial displacement $U_r$ in the proposed solution is substantially superior to the series solution. Figures \ref{Fig5} and \ref{Fig6} show the convergence of the displacements with the truncation limit $N$ under both types of loading. Subsequently, Figures \ref{Fig7} and \ref{Fig8} demonstrate that in the limit as $\phi_0 \to 0$ the displacements due to the distributed loads approach those obtained for the point load. Moreover, the normalized radial displacement, $U_r$, is almost indistinguishable from the point load for a $\phi_0$ as large as 10 degrees. Poisson's ratio of $\nu$=0.4 has been used throughout.

We would also like to investigate how the displacement due to a Hertzian-type load compares with that from the Hertzian contact theory. The dimensionless vertical displacement that we obtain by the methods outlined in this paper has the form
\begin{equation} \label{47}
\begin{aligned}
U_z = U_r \cos \theta - U_\theta \sin \theta   
= (8 \pi GR) \frac {u_z}{F} ,
\end{aligned}
\end{equation}
where $u_z$ is the physical vertical displacement.

The Hertz contact theory \cite{Johnson} is formulated in terms of the radius of the contact area $a$, the displacements directly under the load $\delta$, and the magnitude of the applied load $F$. We need to reformulate these quantities in terms of the contact angle $\phi_0$. The radius of the contact area is simply
\begin{equation} \label{48}
a = R \sin \phi_0 .
\end{equation}
The maximum vertical displacement is related to $a$ in the following manner 
\begin{equation} \label{49}
\begin{aligned}
\delta = \frac{a^2}{R} 
= R \sin^2 \phi_0   
= 2u_z(0) ,
\end{aligned}
\end{equation}
where eq.\ \eqref{48} was used and the last equality arises from the fact that the Hertzian solution presented here is for the contact of two spheres hence we need to half the total displacement. Furthermore, Hertz contact theory tells us that the resultant force $F$ is proportional to $a^3$, or more accurately   
\begin{equation}  \label{554}
F = \frac{4}{3} \Big( \frac{G}{1- \nu} \Big) \frac{a^3}{R}    
= \frac{4}{3} \Big( \frac{G}{1- \nu} \Big) R^2 \sin^3 \phi_0 .
\end{equation}
This allows to rewrite equation \eqref{47} for the dimensionless vertical displacement via Hertz contact theory, denoted as $U_{z}^{H}(0)$. Substituting equations \eqref{49} and \eqref{554} into \eqref{47} yields
\begin{equation} \label{555}
\begin{aligned}
U_{z}^{H}(0) = (8 \pi GR) \frac { \frac{R}{2} \sin^2 \phi_0}{\frac{4}{3} \big( \frac{G}{1- \nu} \big) R^2 \sin^3 \phi_0}   
= \frac{3 \pi (1-\nu)}{\sin \phi_0}  .
\end{aligned}
\end{equation}

\begin{figure} [h]
\centerline{
\includegraphics[width=4.5in, height=3.in]{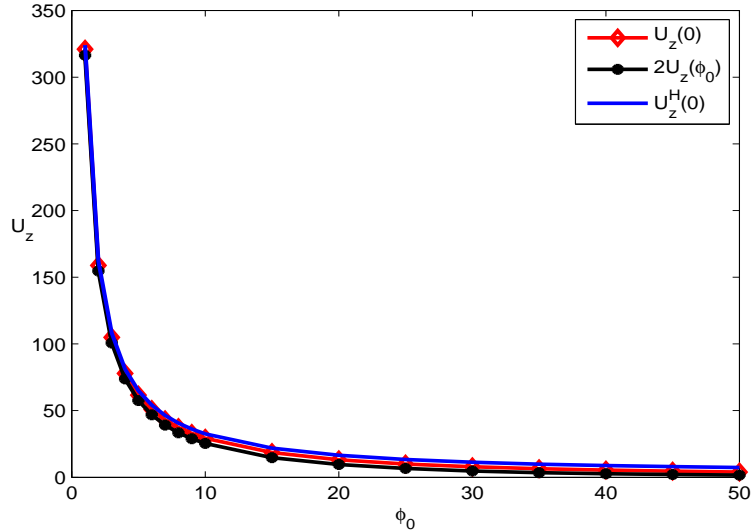}
}
\caption{Comparison of the dimensionless vertical displacement $U_z=(8 \pi GR) \frac {u_z}{F}$ as obtained by the methods in this paper for a Hertzian-type load and that obtained from Hertz contact theory $U_{z}^{H}$ defined in \eqref{555}, as a function of the contact angle $\phi_0$.}
\label{Fig9}
\end{figure}

\begin{figure} [h]
\centerline{
\includegraphics[width=4.5in]{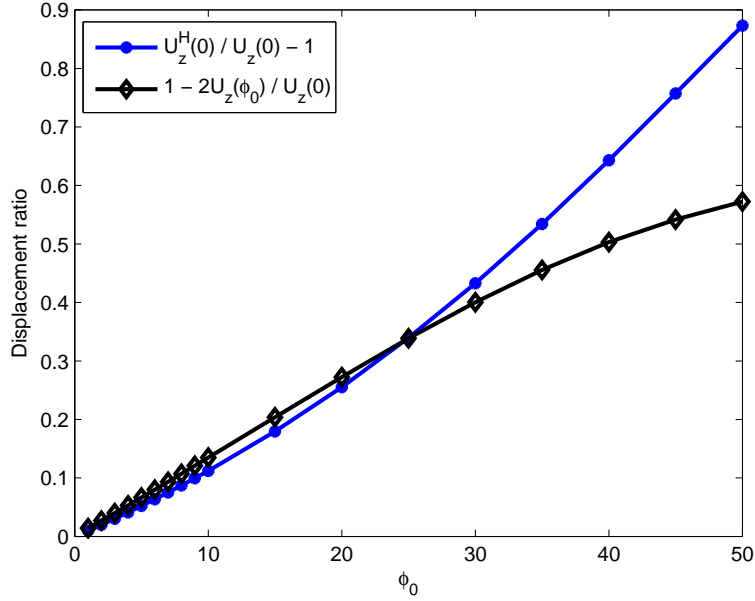}
}
\caption{Normalized difference between the displacements in Figure \ref{Fig9} as a function of the contact angle $\phi_0$.}
\label{Fig10}
\end{figure}

Equation \eqref{555} gives a way to compare the presented solution for the Hertzian-type load to the solution from Hertz contact theory. The numerical results are presented in Figure \ref{Fig9}, which compares the vertical displacements (eq.\ \eqref{47} with eq.\ \eqref{555}) as a function of the contact angle $\phi_0$. Note that along with $U_z(0)$ and $U_{z}^{H}(0)$ we also plot $2U_z(\phi_0)$, which according to Hertz theory should be equal to $U_z(0)$. The normalized difference between the displacements is shown in Figure \ref{Fig10}. As expected, the solutions are close for small contact areas and diverge as this area increases. The same can be said about the relationship between the displacements $U_z(0)$ and $2U_z(\phi_0)$.

\begin{figure} [h]
\centerline{
\includegraphics[width=4.5in]{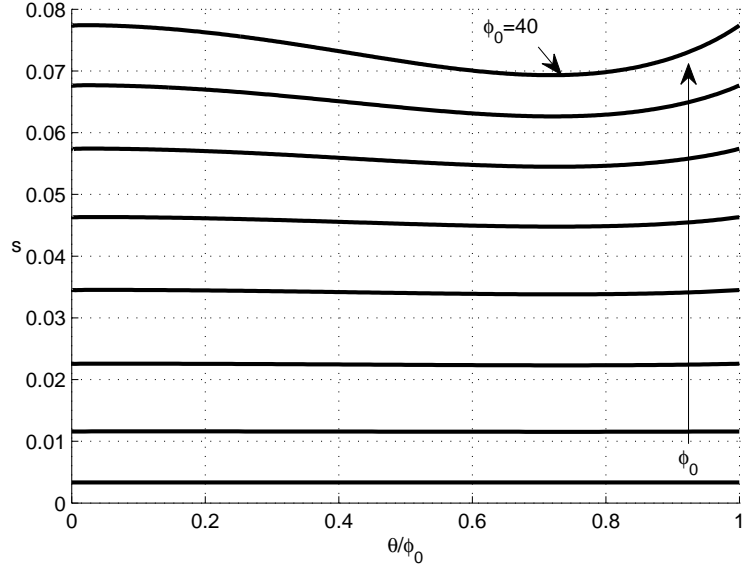}
}
\caption{The flatness of the  area under the load as defined in eq.\ \eqref{g2} as a function of the normalized angle $\theta/\phi_0$. Each curve corresponds to a different contact angle ranging from $\phi_0=5^\circ$ to $40^\circ$ in $5^\circ$ increments.}
\label{Fig11}
\end{figure}

Comparing the maximum displacements $U_z(0)$ with $U_{z}^{H}(0)$ does not tell us anything about the shape of the contact area for a sphere loaded by a Hertzian-type load. Hertz contact theory states that the contact area between two identical spheres is flat, and thus we can describe it using $R(\cos \theta - \cos \phi_0)$. Therefore, we define a function $s(\theta)$ to determine how close is our calculated displacement to the Hertzian solution as
\begin{equation} \label{g1}
s(\theta) = k U_z(\theta) - (\cos \theta - \cos \phi_0),
\end{equation}
where $k$ is a constant determined by enforcing $s(0)=s(\phi_0)$, which results in
\begin{equation} \label{g2}
s(\theta) = \frac{U_z(\theta)}{U_z(0)-U_z(\phi_0)}(1-\cos \phi_0) - (\cos \theta - \cos \phi_0) .
\end{equation}
The function $s(\theta)$ is plotted in Figure \ref{Fig11} for several angles $\phi_0$. These results show that the contact area is flat for small contact angles, but gains curvature for larger angles. According to Hertz theory, for small contact angles $\phi_0$, the function $s(\theta)$ behaves as a constant $s(\theta) \approx \phi_{0}^{2}/2$. The angles shown in Figure \ref{Fig11} are too large to see this behaviour, however, at $\phi_0=5^\circ$ the values are close with $s(\theta)=0.00334$ and $\phi_{0}^{2}/2 = 0.00381$.

\section{Conclusions} \label{sec5}
A compact Green's function for a sphere is presented which uses the fundamental idea of expressing a slowly convergent series with analytical functions and a quickly convergent series. The increased speed of convergence is demonstrated for the point force solution, which is also shown to be consistent with the more general distributed loading in the limit as the contact angle approaches zero. Since the general Green's function contains elliptical integrals, an easy method for dealing with the singularity in the integrand is presented. Comparing the exact displacement due to a Hertzian-type distributed load to the displacement given by Hertz contact theory we conclude that the Hertz contact theory gives accurate results for contact angles up to about $10^\circ$, with a steadily increasing error. For larger contact angles, Hertz theory overestimates the displacements and cannot account for the shape of the contact area. This is to say that the stress distribution assumed in Hertz theory results in a curved contact surface for larger contact angles.

\appendix
\section*{Appendix A: Legendre polynomial formulas}

The orthogonality and completeness relations for the Legendre functions are  
\begin{subequations} \label{4}
\begin{align}
 \frac 1{2}{(2n+1)}\int\limits_0^\pi P_m(\cos \theta ) P_n(\cos \theta ) \, \sin  \theta \, d \theta &= 
 \delta_{mn}, 
\\
\sum\limits_{n=0}^\infty   \frac 1{2} (2n+1)  P_n(\cos \theta ) P_n(\cos \phi ) &= 
\frac{\delta (\theta - \phi)}{\sin \phi} , 
\end{align} 
\end{subequations}

Starting with the definition for $P_n(x)$, 
\begin{equation} \label{51}
P_n(\cos\theta ) 
= \frac 1{\pi} \int_0^\pi (\cos\theta +i\sin \theta \cos\alpha )^n\, d \alpha ,
\end{equation}
and using $\sum\limits_{n=0}^\infty z^n = \frac 1{1-z}$, $|z|< 1$, 
the well known generating function follows 
\begin{equation} \label{8}
 \sum\limits_{n=0}^\infty  {t}^n  P_n(\cos \theta )
= \frac 1{\sqrt{ 1+t^2 -2t \cos\theta} } , \ \  |t|\le 1 . 
\end{equation}
Integrating the identity  \eqref{8}
with respect to $t$ 
implies 
\begin{align} \label{9}
 &\sum\limits_{n=0}^\infty t^{n+1}   \frac{ P_n(\cos \theta )}{n+1}
 = \sinh^{-1} ( \cot \theta   ) 
+ \sinh^{-1}\bigg( \frac{t- \cos \theta }{\sin\theta} 
\bigg) 
\notag \\
& \quad =
\log \big( 1+ \csc \frac \theta 2 \big) 
- \sinh^{-1} ( \tan \frac{\theta}2   )  +  \sinh^{-1}\Big( \frac{t- \cos \theta }{\sin\theta}  \Big)
, \ \ |t|\le 1 , \ \ 0\le \theta \le \pi. 
\end{align}
Taking the limit as $t \to 1$ yields \eqref{22a}. 
$S_1(\theta) $ of \eqref{22b} follows from  a similar result \cite[eq.\ 5.10.1.4]{Prudnikov2}, while  $S_2(\theta)$  of \eqref{22c} follows from the recurrence relation 
\begin{equation} \label{21}
(n+1) P_{n+1} (x) - (2n+1) x P_{n } (x) + n P_{n-1} (x) = 0, 
\end{equation}
after  dividing by   $n$  and summing from $n=1$   to $\infty$   ($S_2$ agrees with 
\cite[eq.\ 5.10.1.6]{Prudnikov2}).  
The recurrence relation can be used to then find 
$S_j  (\theta ) $ for $j =  3$, $4, \ldots $.


Series of products of Legendre functions given by eqs.\ 6.11.3.1 and 6.11.3.2 of 
\cite{Brychkov} 
\begin{equation} \label{1-2}
\begin{aligned}
\sum\limits_{n=1}^\infty   \frac{2n+1}{n(n+1)}   P_n(x ) P_n(y )
& = - 1 - \log \tfrac{ (1-x)(1+y)}{4} 	, 
\\
\sum\limits_{n=1}^\infty   \frac{2n+1}{n^2(n+1)^2}   P_n(x ) P_n(y )
& = 1 - \log \tfrac{ 1+y}{2} \log \tfrac{ (1-x)(1+y)}{4}  
+ Li_2 \big( \tfrac {1+x}2 \big) - Li_2 \big( \tfrac {1+y}2 \big) ,
\end{aligned}
\end{equation}
for $-1 \le x < y \le 1$.     Equation \eqref{1-2}$_1$ can be derived by operating on both sides by the Legendre differential operator $L_x = \frac{d}{dx}(1-x^2)\frac{d}{dx}$, and using the eigenvalue property
$L_x P_n(x) = -n(n+1) P_n(x)$ to arrive at \eqref{4b} (for $x<y$).  At the same time, the constants in 
the right member of  \eqref{1-2}$_1$ follow by considering the formula for $x=0$, $y=1$ in which case the sum on the left can be found.   	
Equation \eqref{1-2}$_1$ gives $S_0(\theta,\phi)+S_1(\theta,\phi)$ by noting that 
$\frac 1n + \frac 1{n+1} = \frac{2n+1}{n(n+1)}$. 

The following is a simple consequence of Legendre's addition formula \cite[eq.\ 3.19]{Martin06}
\begin{equation} \label{2-2}
P_n(\cos\theta )P_n(\cos\phi ) 
= \frac 1{\pi} \int_0^\pi 
P_n(\cos\theta \cos\phi- \sin\theta \sin\phi \cos\alpha) 
\, d \alpha .
\end{equation}
Multiply both sides of \eqref{2-2} by $t^n$ and sum, implies, using eq.\ \eqref{8}, the identity 
\cite[eq.\ 5.10.2.1]{Prudnikov2} for $|t| <1$, 
\begin{equation} \label{1-4}
\sum\limits_{n=0}^\infty    t^{n}\, P_n(\cos\theta )P_n(\cos\phi ) 
= \frac{4}{\pi ( u_++u_-)}\, 
K \Big(\frac{ u_+ -u_-}{ u_++u_-}
\Big) , 
\\ 
u_\pm = \sqrt{
1-2 t \cos(\theta \pm \phi) + t^2}  .
\end{equation}

\section*{Appendix B: Analytical functions and their derivatives}

We require the derivatives with respect to $\theta$ of the  functions defined in eq.\ \eqref{22}. They are
\begin{equation} \label{B2}
\begin{aligned}
\frac{dS_0 (\theta )}{d\theta} &=  \frac{\sin\frac \theta 2  - 1}{\sin \theta} , 
\\
\frac{dS_1 (\theta )}{d\theta} &= - \frac{dS_0 (\theta )}{d\theta}   - \cot \frac \theta 2 , 
\\
\frac{dS_2 (\theta )}{d\theta} &= \frac{dS_1 (\theta )}{d\theta}\cos\theta -  S_1 (\theta ) \sin\theta  + \cos\frac \theta 2 \Big( 2 \sin \frac \theta 2 - 1 \Big)  .
\end{aligned}
\end{equation}
Similarly, the analytical function used to find $u_\theta$ (eq.\ \eqref{4=5}) in section \ref{sec4} is
\begin{equation} \label{B3}
S_{01}(\theta,\phi)  =  S_0(\theta,\phi)+S_1(\theta,\phi) = 
\begin{cases}
- 2\log  	\sin \frac {\theta}2 \cos \frac {\phi}2  , & \theta > \phi, 
\\
- 2\log  	\cos \frac {\theta}2 \sin \frac {\phi}2  , & \theta < \phi .
\end{cases} 
\end{equation}
The derivative of $S_{01}(\theta,\phi)$ is
\begin{equation} 
\begin{aligned}
\frac{\partial S_{01}(\theta,\phi)}{\partial\theta}  = 
\begin{cases} 
-\cot\frac{\theta}{2}  , & \theta > \phi, 
\\
\tan \frac{\theta}{2}  , & \theta < \phi .
\end{cases}
\end{aligned}
\end{equation}



\end{document}